\documentclass[aps,preprint]{revtex4-1}
\usepackage{graphicx}
\usepackage{amsmath}
\usepackage{makeidx}
\usepackage{amsfonts}
\usepackage{amssymb}
\usepackage{dsfont}
\usepackage{color,soul}
\usepackage[autostyle]{csquotes}

\begin{document}

\title{Statistical Physics of Medical Diagnostics: Study of a Probabilistic Model}

\author{Alireza Mashaghi$^{a,b,*}$, Abolfazl Ramezanpour$^{a,c}$}
\affiliation{$^a$Leiden Academic Centre for Drug Research, Faculty of Mathematics and Natural Sciences, Leiden University, Leiden, The Netherlands}
\affiliation{$^b$Harvard Medical School, Harvard University, Boston, Massachusetts, USA}
\affiliation{$^c$Department of Physics, University of Neyshabur, Neyshabur, Iran}
\affiliation{$*$a.mashaghi.tabari@lacdr.leidenuniv.nl}

\date{\today}

\begin{abstract}
We study a diagnostic strategy which is based on the anticipation of the diagnostic process by simulation of the dynamical process starting from the initial findings. We show that such a strategy could result in more accurate diagnoses compared to a strategy that is solely based on the direct implications of the initial observations. We demonstrate this by employing the mean-field approximation of statistical physics to compute the posterior disease probabilities for a given subset of observed signs (symptoms) in a probabilistic model of signs and diseases. A Monte Carlo optimization algorithm is then used to maximize an objective function of the sequence of observations, which favors the more decisive observations resulting in more polarized disease probabilities. We see how the observed signs change the nature of the macroscopic (Gibbs) states of the sign and disease probability distributions. The structure of these macroscopic states in the configuration space of the variables affects the quality of any approximate inference algorithm (so the diagnostic performance) which tries to estimate the sign/disease marginal probabilities. In particular, we find that the simulation (or extrapolation) of the diagnostic process is helpful when the disease landscape is not trivial and the system undergoes a phase transition to an ordered phase.           
\end{abstract}

%\pacs{} 
\maketitle

\section{Introduction}\label{S0}
Statistical physics has been widely used to extract macroscopic properties of a wide range of systems from their microscopic interaction models, yet it has not been employed to medical diagnostics. Given an initial subset of observed signs (symptoms, clinical and laboratory findings) with some prior knowledge about the patient (or a complex system like a biological cell), a diagnosis problem simply asks for the most probable diseases (or macrostates with specific phenotypes) \cite{Ledley,MillerDDSS,Current-2016}. An efficient and accurate diagnostic procedure is important specially in the early stages of diseases, where the number and quality of medical evidences are often insufficient to reach a definite diagnosis. Here, we use approximate inference and optimization algorithms of statistical physics \cite{HR-book-2006,MM-book-2009} to show that a simulation (extrapolation) of the diagnostic process (without doing any real observation) could be helpful as a heuristic strategy in the study of diagnosis problems.

A diagnosis problem usually starts with a (probabilistic) model of (well-defined) sign and disease variables which describes the (statistical) dependencies of the variables; such an \enquote{effective} model of the signs and diseases may come from a microscopic model of the system (human body or biological cell) with emergent macroscopic behaviors that are interpreted as diseases. Various modeling frameworks have been developed and used in medical diagnostics: (i) probabilistic models and belief networks, (ii) neural networks and machine learning methods, and (iii) a complex network approach to the problem.

Bayesian belief networks provide a probabilistic framework to study the sign-disease dependencies \cite{Spiegelhalter,internist-1,Heckerman-1,Miller94,Nikovski-ieee-2000}. The belief networks are represented by tables of conditional probabilities that show how the state of a variable in an acyclic directed graph depends on the state of the parent variables. The above information along with a few simplifying assumptions then are used to infer the marginal sign and disease probabilities for a given set of findings \cite{internist-1}. Another approach is to use an artificial neural network to represent the complex relationships of the sign/disease variables \cite{B-nc-1990,P-ieee-1998,Khan-nm-2001}. The model parameters here are obtained in a learning process using the machine learning techniques \cite{Murphy-book}. Finally, in a network approach to the problem, one constructs a (weighted) symptom-disease network with connections relating the signs to the diseases. This network along with other complementary data, e.g., gene-disease, RNA-disease, protein-disease, metabolite-disease and disease-disease networks, are then utilized as information resources by a diagnostic algorithm \cite{Goh-pnas-2007,Barabasi-nr-2011,G-gm-2014,Sun-bmc-2014,Liu-sr-2015,Suratanee-2015}.

In addition to the model, an efficient inference algorithm is needed to estimate the marginal sign and disease probabilities \cite{cooper-ai-1990,Jordan}. When the number of initially observed signs is too small to make a diagnosis, we need to suggest a number of new medical tests to know the value of several other relevant signs. For this aim, we need an appropriate objective function and optimization algorithm to choose the more informative signs, which can lead us to the right diagnosis with a smaller number of observations. In Refs. \cite{paper-1,paper-2} we proposed probabilistic models of signs and diseases which can systematically take into account the effects of different types of sign-disease, disease-disease, and sign-sign interactions; the models are indeed graphical models of the sign and disease variables with a number of interaction factors, each one connecting a small subset of disease variables to the associated sign variables \cite{Jordan}. We introduced approximate inference and optimization algorithms to deal with such probabilistic models, and studied the effects of the model structure and the objective function on the performances of the diagnostic algorithms.

The models we consider are natural generalizations of the simpler probabilistic models studied in previous works \cite{internist-1,Heckerman-1,Nikovski-ieee-2000}, which usually assume that only one disease is behind the findings (exclusive diseases assumption), or, the diseases act independently on the signs (causal independence assumption). Moreover, for computational simplicity, it is usually assumed that there is no disease-disease and sign-sign interactions. In Ref. \cite{paper-1}, we showed that such interactions can significantly improve the accuracy of diagnosis without resorting to the exclusive diseases or the causal independence assumption.

In this paper, we elaborate more on the nature and behavior of the macroscopic states of the probabilistic models we introduced in the above studies. We employ the mean-field approximation to study the possible changes in the (macroscopic) state of the system as the number of observed signs increases, and to estimate the sign and disease marginal probabilities \cite{MF-book-2001}. For the objective function, we choose a function that favors the more polarizing tests, which result in disease probabilities that are closer to zero or one \cite{paper-1}. This could be useful especially when the gap between the most probable diseases and the other diseases is small. Moreover, this objective function is easier to compute than a maximum-likelihood function that is typically taken in these problems. Starting from an initial set of observed signs, we use an approximate (Monte Carlo) optimization algorithm to find a sequence of candidate observations that maximizes the above objective function \cite{paper-2}. However, instead of the true value of the \enquote{observed} sign at each step, we assume the outcome is given by the most probable value of the sign obtained from the model by the approximate inference algorithm. We show that this strategy is able to improve the quality of diagnosis compared to the case that is merely based on the direct implications of the initial findings. Interestingly, the improvement is observed for nontrivial cases when the system undergoes a phase transition to an ordered phase; i.e., where the effect of observed signs can propagate in the system to influence the state of the other signs (for a similar phenomenon see \cite{MM-jstat-2006}).

\section{Main definitions}\label{S1}
The microscopic state of the system (patient or cell) is identified by the sign values $\mathbf{S}=\{S_i: i=1,\dots,N_S\}$, where for simplicity we work with binary sings $S_i=\pm 1$. The probability of being in state $\mathbf{S}$ is given by $P(\mathbf{S})$. The probability $P(\mathbf{S})$ (the model) is parametrized by a set of couplings $\mathbf{K}(t)$, which in general depend on real time $t$. The conditional probability of the unobserved signs depends on the subset of the observed signs $\mathbf{O}(t)=\{j=1\dots,N_O(t)\}$ with the values denoted by $\mathbf{S}^o(t)=\{S_j:j \in \mathbf{O}(t)\}$. 

The macroscopic states (or phenotypes, or diseases) of the system (for large number of signs $N_S\to \infty$) can be identified by the Gibbs states of $P(\mathbf{S})$ \cite{gibbs-pnas-2007}. We label these macroscopic states with $\mathbf{D}$, for diseases, with $\mathbf{D}=\mathbf{0}$ representing the healthy state. The average of an unobserved sign in state $\mathbf{D}$ is denoted by $\langle S_i\rangle_{\mathbf{D}}$. A pure Gibbs state is characterized by the clustering property; i.e., $\langle S_iS_j\rangle_{\mathbf{D}}-\langle S_i\rangle_{\mathbf{D}}\langle S_j\rangle_{\mathbf{D}}$ goes exponentially to zero by the distance of sign nodes $i$ and $j$ in the interaction graph of the sign variables induced by $P(\mathbf{S})$. A mixed state is composed of more than one pure states. In this way, the state of a disease pattern $\mathbf{D}$ or a cluster of similar disease patterns are represented by the statistical properties of the sign variables in the associated pure or mixed Gibbs states.

We start by asking several interesting questions:
\begin{itemize}

\item  What is $P(\mathbf{S})$ and how does it (or the couplings $\mathbf{K}$) change with time? Here we need a dynamical model of the system to study the stochastic evolution of the sign variables. In the following, we shall assume some reasonable structures for the model and leave this problem for future studies. Instead of going from the model $P(\mathbf{S})$ to the macroscopic states, we start from the diseases $\mathbf{D}$ and obtain the model from the joint probability of the sign and disease variables $P(\mathbf{S};\mathbf{D})=P(\mathbf{S}|\mathbf{D})P_0(\mathbf{D})$. Then, the model is obtained by summing over the disease variables $P(\mathbf{S})=\sum_{\mathbf{D}}P(\mathbf{S}|\mathbf{D})P_0(\mathbf{D})$. The conditional probability $P(\mathbf{S}|\mathbf{D})$ can be a decreasing function of the distance of $\mathbf{S}$ and a reference sign configuration $\mathbf{S}(\mathbf{D})$. Here the $S_i(\mathbf{D})$ represent the most probable symptoms of disease $\mathbf{D}$. These models could be useful (in the absence of the realistic models) as benchmarks in the study of a diagnosis problem. 
   
\item  Do we see a significant change of behavior with time in $P(\mathbf{S})$? For example, from weak sign correlations to a regime of strong correlations. Typically, we encounter strong correlations close to a phase transition from one macroscopic state to another state. As we will see, even simple (but plausible) models of signs and diseases can exhibit both continuous and discontinuous phase transitions as the strength of the sign and disease interactions are varied. In particular, the phase transition can be induced by increasing the number of observed signs for given strength of the interactions.   

\item  How does the structure of $P(\mathbf{S})$ affect the diagnosis? Here we need an efficient and approximate inference algorithm to compute the sign and disease probabilities. It is easy to obtain very good estimations of these marginal probabilities as long as there is only one macroscopic (pure) state, or there are a number of symmetry-related states. Otherwise, the above algorithms will not converge or will need a very large computation time to provide a fair sampling of the probability distribution. We will see how the convergence and quality of an approximate inference algorithm which is based on the mean-field approximation affect the diagnostic performance. 

\end{itemize}

Consider a set of $N_D$ binary variables $\mathbf{D}=\{D_{a}=0,1: a=1,\dots,N_D\}$, where $D_{a}=0,1$ shows the absence or presence of disease $a$. Each disease is assigned a positive weight $W_a$, to take into account the significance of diseases. In the following we assume the $W_a$ are uniformly distributed in $(0,1)$. The joint probability distribution of the sign and disease variables (i.e., the model) is identified by $P(\mathbf{S};\mathbf{D})=P(\mathbf{S}|\mathbf{D})P_0(\mathbf{D})$. Here $P_0(\mathbf{D})$ is the prior probability distribution of diseases, which could depend on the patient's characteristics such as gender and age and disease properties such as duration of a disease, mortality rate and transmission rate among others. In the following, we shall assume the prior probability is factorized as $P_0(\mathbf{D}) \propto \exp(\sum_a K_a^0D_a)$. The parameters $K_a^0$ can also be used to control the expected number of present diseases.

Let $P_{true}(\mathbf{S}|\mathbf{D})$ to be the true (or empirically estimated) probability distribution of the sign variables given the disease hypothesis $\mathbf{D}$. In practice, we may have access only to a small subset of marginal probabilities of the above probability distribution. For instance, suppose we are given the sign probabilities $P_{true}(S_i|\mathrm{nodisease})$, $P_{true}(S_i,S_j|\mathrm{only}D_a)$, and $P_{true}(S_i,S_j|\mathrm{only}D_a,D_b)$ conditioned on the presence of no diseases, the presence of only one disease, and the presence of only two diseases, respectively. Using the maximum entropy principle, for the conditional probability distribution of the signs we take
\begin{align}\label{PSD}
P(\mathbf{S}|\mathbf{D})=\frac{1}{Z(\mathbf{D})}\phi_0(\mathbf{S}) \times \prod_a \phi_a(\mathbf{S}|D_a)\times  \prod_{a < b} \phi_{ab}(\mathbf{S}|D_a,D_b),
\end{align}
where the partition function $Z(\mathbf{D})$ is obtained from normalization. 

\begin{figure}
\includegraphics[width=8cm]{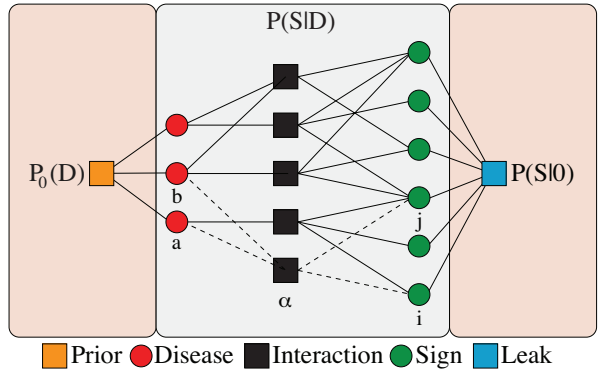} 
\caption{The interaction graph of disease variables (left circles) and sign variables (right circles) related by $M_a$ one-disease and $M_{ab}$ two-disease interaction factors (middle squares) in addition to interactions induced by the leak probability (right square) and the prior probability of diseases (left square). In general, an interaction factor $\alpha=a,ab$ is connected to $k_{\alpha}$ signs and $l_{\alpha}$ diseases \cite{paper-1}.}\label{dsf}
\end{figure}

The disease interaction factors ($\phi_0, \phi_a, \phi_{ab}$) can in general be parametrized by the couplings of all the possible multi-sign interactions. As customary in maximum entropy modeling, assuming an exponential family, the parameters
sufficient to describe the above marginal probabilities are involved in the one-sign terms ($K_i^0S_i$), the one-disease-one-sign interactions ($K_i^aD_aS_i$), the one-disease-two-sign interactions ($K_{ij}^aD_aS_iS_j$), the two-disease-one-sign interactions ($K_i^{ab}D_aD_bS_i$), and finally the two-disease-two-sign interactions ($K_{ij}^{ab}D_aD_bS_iS_j$). More precisely, the disease interaction factors are given by 
\begin{align}
\phi_0(\mathbf{S}) &= e^{\sum_i K_i^0 S_i},\\
\phi_{a}(\mathbf{S}|D_a) &= e^{D_a[\sum_i K_i^a S_i+\sum_{i<j} K_{ij}^a S_iS_j]},\\
\phi_{ab}(\mathbf{S}|D_a,D_b) &= e^{D_aD_b[\sum_i K_i^{ab} S_i+\sum_{i<j} K_{ij}^{ab} S_iS_j]}.
\end{align}
Figure \ref{dsf} shows the interaction graph of the sign and disease variables related by the above interaction factors. We use $M_a,k_a$ and $M_{ab},k_{ab}$ for the number and connectivity of one-disease and two-disease interaction factors, respectively.

In principle, the information provided by the marginal probabilities of the true (or empirical) probability distribution is sufficient to determine the model parameters \cite{KR-nc-1998,T-pre-1998,T-jstat-2012,inverse-2017}. 
Note that $\phi_0(\mathbf{S})$ is responsible for the probability of observing $\mathbf{S}$ in the absence of any diseases, where the most probable value is $S_i=-1$. It is reasonable to assume that in the healthy case each sign takes the positive value with a small probability independently of the other sign values.  

The joint probability distribution of the sign and disease variables can be rewritten as, $P(\mathbf{S};\mathbf{D}) \propto \exp(-\mathcal{H}(\mathbf{S};\mathbf{D}))$, where the energy function reads as follows
\begin{align}
\mathcal{H}(\mathbf{S};\mathbf{D}) = -\sum_i h_i(\mathbf{D}) S_i-\sum_{i<j}J_{ij}(\mathbf{D})S_iS_j
+\ln Z(\mathbf{D})-\ln P_0(\mathbf{D}).
\end{align}
Here, the partition function and the new couplings are:
\begin{align}
Z(\mathbf{D}) &=\sum_{\mathbf{S}} e^{\sum_i h_i(\mathbf{D}) S_i+\sum_{i<j}J_{ij}(\mathbf{D})S_iS_j},\\
h_i(\mathbf{D}) &= K_i^0+\sum_{a}K_i^{a}D_{a}+\sum_{a<b}K_i^{ab}D_{a}D_{b},\\
J_{ij}(\mathbf{D}) &= \sum_{a}K_{ij}^{a}D_{a}+\sum_{a<b}K_{ij}^{ab}D_{a}D_{b}.
\end{align}
From the above model, we can extract simpler models depending one the maximum number of disease and sign variables that are involved in the interactions; for instance, we could have the D1S1 (one-disease-one-sign), D1S2 (one-disease-two-sign), D2S1 (two-disease-one-sign), and D2S2 (two-disease-two-sign) models.

In the following, we consider only the $D1S1$ and $D2S1$ models, where we can exactly compute the partition function 
$Z(\mathbf{D})=\prod_i \left( 2\cosh h_i(\mathbf{D})\right)$.
For these models, we can also exactly compute the model parameters given the true marginal probabilities,
\begin{align}
K_i^0 &=\frac{1}{2}\ln\left(\frac{P_{true}(S_i=+1|\mathrm{nodisease})}{P_{true}(S_i=-1|\mathrm{nodisease})}\right),\\
K_i^{a} &=\frac{1}{2}\ln\left(\frac{P_{true}(S_i=+1|\mathrm{only} D_a)}{P_{true}(S_i=-1|\mathrm{only} D_a)}\right)-K_i^0,\\
K_i^{ab} &=\frac{1}{2}\ln\left(\frac{P_{true}(S_i=+1|\mathrm{only} D_a,D_b)}{P_{true}(S_i=-1|\mathrm{only} D_a,D_b)}\right)-K_i^0-K_i^a-K_i^b.
\end{align}

For a given subset $\mathbf{O}$ of observed signs with values $\mathbf{S}^o$, the disease probabilities are obtained from 
\begin{align}
P(D_a=1|\mathbf{S}^o)=\frac{1}{\mathcal{Z}(\mathbf{S}^o)}\sum_{\mathbf{D}} D_a e^{-\mathcal{H}(\mathbf{D}|\mathbf{S}^o)},
\end{align}
where $\mathcal{H}(\mathbf{D}|\mathbf{S}^o) = -\log \mathcal{L}(\mathbf{D}|\mathbf{S}^o)$ is the log-likelihood function
\begin{align}
\mathcal{H}(\mathbf{D}|\mathbf{S}^o)=  -\sum_a K_a^0D_a-
\sum_{i\in \mathbf{O}}S_i^oh_i(\mathbf{D})+\sum_{i\in \mathbf{O}}\ln \left(2\cosh h_i(\mathbf{D}) \right),
\end{align}
and $\mathcal{Z}(\mathbf{S}^o)= \sum_{\mathbf{D}} \exp(-\mathcal{H}(\mathbf{D}|\mathbf{S}^o))$ is another normalization constant.
As before, the prior probability distribution is $P_0(\mathbf{D}) \propto \exp(\sum_a K_a^0 D_a)$.
It is easy to show that the marginal probability of an unobserved sign is given by:
\begin{align}
P(S_i=1|\mathbf{S}^o)=\frac{1}{\mathcal{Z}(\mathbf{S}^o)}\sum_{\mathbf{D}} \left(\frac{1+\tanh h_i(\mathbf{D})}{2}\right) e^{-\mathcal{H}(\mathbf{D}|\mathbf{S}^o)}.
\end{align}
The approximate equations for the D1S2 and D2S2 models can be found in \cite{paper-1}.

\section{The homogeneous fully-connected models}\label{S2}
The direct problem of inferring the marginal sign/disease probabilities from the above models can be solved exactly as long as the model parameters do not depend on the sign or disease labels.
The thermodynamic limit here is defined by the limit $N_D,N_S,N_O \to \infty$ such that $\gamma= N_D/N_S$ and $n_o=N_O/N_S$ remain finite. To provide some order of magnitude, it is useful to mention that in Internist (a probabilistic model for internal diseases \cite{internist-1}) the number of diseases is about $500$ and the number of associated signs is around $4000$. In addition, we need to scale the model parameters as $K_{ij}^a=\kappa_{ij}^a/(N_SN_D), K_{ij}^{ab}=\kappa_{ij}^{ab}/(N_SN_D^2)$, $K_i^0=\kappa_i^0, K_i^a=\kappa_i^a/N_D, K_i^{ab}=\kappa_i^{ab}/N_D^2$, and $K_a^0=\kappa_a^0$; the scaling ensures that the energy function is extensive (proportional to $N_S$).

\begin{figure} 
\includegraphics[width=10.5 cm]{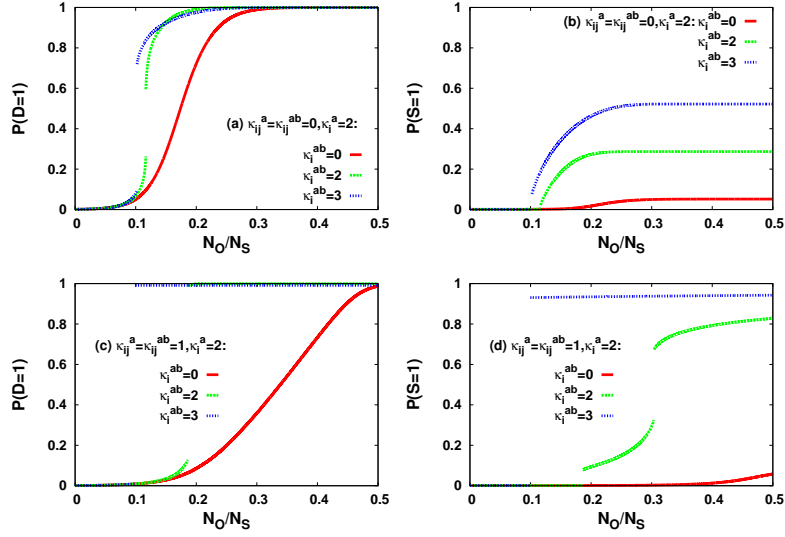} 
\caption{The sign and disease probabilities in the homogenuous fully-connected models vs the fraction of observed signs $N_O/N_S$.  We assume that all the observed signs are positive.  The prior disease probabilities and the leak sign probabilities are $P_0(D_a=1)=P(S_i=+1|\mathbf{0})=0.001$. Panels (a),(b) show the probability of observing a positive sign $P(S=1)$ and the probability of having a disease $P(D=1)$ for the D1S1 and D2S1 models ($K_{ij}^a=K_{ij}^{ab}=0$). Panels (c),(d) display the above probabilities for the D1S2 and D2S2 models ($K_{ij}^a=K_{ij}^{ab}=1$).}\label{HMF-no}
\end{figure}

The sign and disease probabilities are obtained by minimizing the following free energy with respect to $x=P(D=1)$ and $y_u=P(S=1)$ (for an unobserved sign), 
\begin{multline}
f(x,y)=-\gamma \mathcal{S}(x)-(1-n_o)\mathcal{S}(\frac{1+y_u}{2})+\mathcal{S}(\frac{1+z(x)}{2})\\
-h(x)(y-z(x))-\frac{1}{2}J(x)(y^2-z^2(x))-\gamma \kappa_a^0 x.
\end{multline}
The value of the observed signs enters in $y=n_o y_o+(1-n_o)y_u$ with $y_o=(\sum_{i\in \mathbf{O}}S_i^o)/N_O$, and $y_u=(\sum_{i\notin \mathbf{O}}S_i)/(N_S-N_O)$. Here $z$ is the solution to $z=\tanh(h(x)+J(x) z)$, and $\mathcal{S}(x)=-x\log x-(1-x)\log(1-x)$ is the Gibbs-Shannon entropy function. Moreover, the effective field $h(x) =\kappa_i^0+\kappa_i^a x+\frac{1}{2}\kappa_i^{ab}x^2$ and the coupling $J(x) =\kappa_{ij}^a x+\frac{1}{2}\kappa_{ij}^{ab}x^2$ (see Appendix \ref{app-1} for the derivations). Each local or global minimum of the free energy can be considered as a macroscopic state of the system.  Figure \ref{HMF-no} shows how the sign and disease probabilities change with the number of observations, when all the observed signs have a positive value (see also Fig. \ref{HMF-fx} in Appendix \ref{app-1}). As the figures show, a new macroscopic state can appear continuously or discontinuously depending on the value of the model parameters.

\section{The inhomogeneous models: Mean-field approximation}\label{S3}
In this section, we find an estimation of the sign and disease probabilities for arbitrary couplings $\mathbf{K}$. To this end, we write $D_a=\langle D_a \rangle+\delta D_a$ and $S_i=\langle S_i \rangle+\delta S_i$ where the $\delta D_a=D_a-\langle D_a \rangle$ and $\delta S_i=S_i-\langle S_i \rangle$ are small deviations from the mean values. The mean-field (MF) approximation here is obtained by neglecting the second order deviations in a Taylor expansion around the mean values \cite{MF-book-2001}. In the following, we shall restrict ourselves to the D1S1 and D2S1 models, where the normalization function $Z(\mathbf{D})$ can be computed exactly; for the D1S2 and D2S2 models we need also to compute this function within the MF approximation (see Appendix \ref{app-2}).

In this way, the MF approximation for the sign and disease probabilities are obtained by solving the self-consistency equations $x_a =\exp(h_a(\mathbf{x}))/(1+\exp(h_a(\mathbf{x})))$, with $P(D_a=1)=x_a$ and $P(S_i=1)=(1+\tanh h_i(\mathbf{x}))/2$. Here, the effective fields experienced by the sign and disease variables are given by
\begin{align}
h_i(\mathbf{x}) &= K_i^0+ \sum_a K_i^a x_a+\sum_{a<b}K_{i}^{ab}x_ax_b,\\
h_a(\mathbf{x}) &= K_a^0+\sum_{i\in \mathbf{O}}[S_i^o-\tanh(h_i(\mathbf{x}))](K_i^a+\sum_{b\neq a}K_i^{ab}x_b).
\end{align}
The equations are solved by iteration starting from random initial values for the $x_a$. The time complexity of this algorithm is of order $N_ON_D^3$ in a fully-connected model. The fixed points of these equations are considered as the macroscopic states of the system (Gibbs states). As long as there is only one macroscopic state, the iteration algorithm converges easily to the single fixed point of the equations. Non-convergence of the iteration algorithm is a signature of the presence of more than one fixed point.

To check the performances of the algorithms, we shall assume that the true model is given by an exponential probability distribution $P_{true}(\mathbf{S}|\mathbf{D}) \propto \exp(-\beta H(\mathbf{S};\mathbf{S}(\mathbf{D})))$. Here $\mathbf{S}(\mathbf{D})$ gives the most probable symptoms of disease pattern $\mathbf{D}$, and $H(\mathbf{S};\mathbf{S}(\mathbf{D}))$ is the Hamming distance (number of different elements) of the two sign configurations. Moreover, $\beta$ is a positive parameter that controls the structure of the true model around the symptoms $\mathbf{S}(\mathbf{D})$; the diseases are more clearly distinguished for larger values of $\beta$. We assume that each element $S_i(\mathbf{D})$ (for $i=1\dots,N_S$) takes the positive and negative values with equal probability, except for the healthy case ($\mathbf{D}=\mathbf{0}$) where all the elements are negative. Given the true model, we use the true marginal probabilities to construct e.g. the D2S1 model. 

\begin{figure}
\includegraphics[width=16cm]{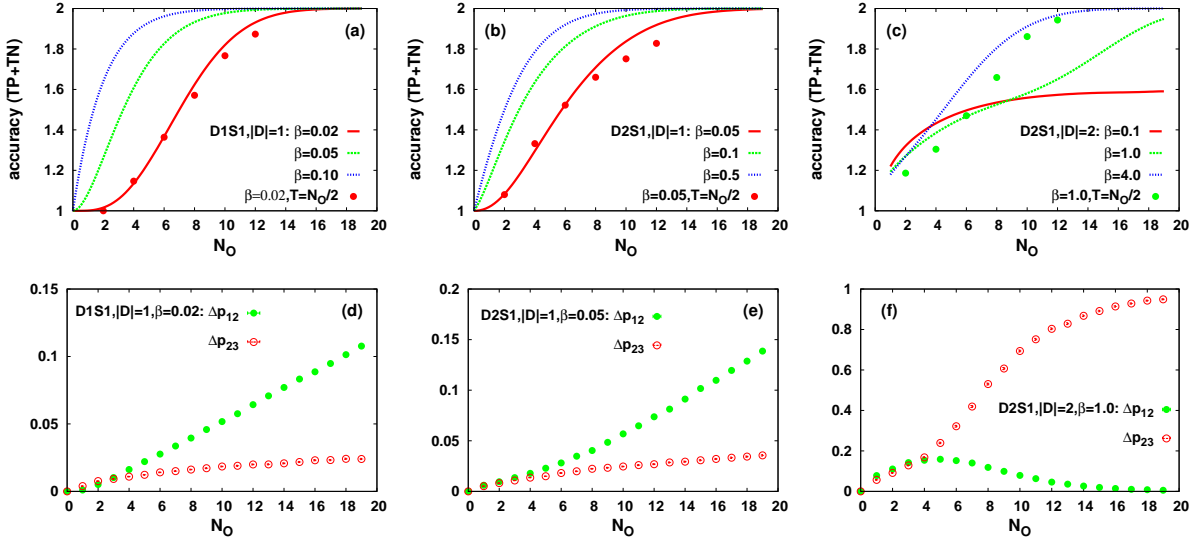} 
\caption{Comparing the accuracy (true positive plus true negative) of the diagnosis (panels a,b,c) and the relative gaps (panels d,e,f) in the disease probabiltiies (sorted by magnitude) $\Delta p_{12}=(P_1-P_2)/P_1$ and $\Delta p_{23}=(P_2-P_3)/P_2$ for the D1S1 and D2S1 models using an exhastive inference algorithm. We consider the cases in which only one or two diseases are present $(|\mathbf{D}|=1,2)$. The prior disease probabilities are chosen such that $N_DP_0(D_a=1)=|\mathbf{D}|$. The model parameters are obtained from the exponential true model for different values of $\beta$. The filled circles in the top panels show the results after a simulation process of $T=N_O/2$ steps, where $N_O$ is the initial number of the observed signs with known true values. For the model structure we take a fully-connected graph of $N_D=5, N_S=20$ variables with $M_a=5, M_{ab}=10$ interaction factors, and connectivities $k_a=k_{ab}=20$. The data are results of $2000$ independent realizations of the problem.}\label{EX}
\end{figure}

\begin{figure}
\includegraphics[width=16cm]{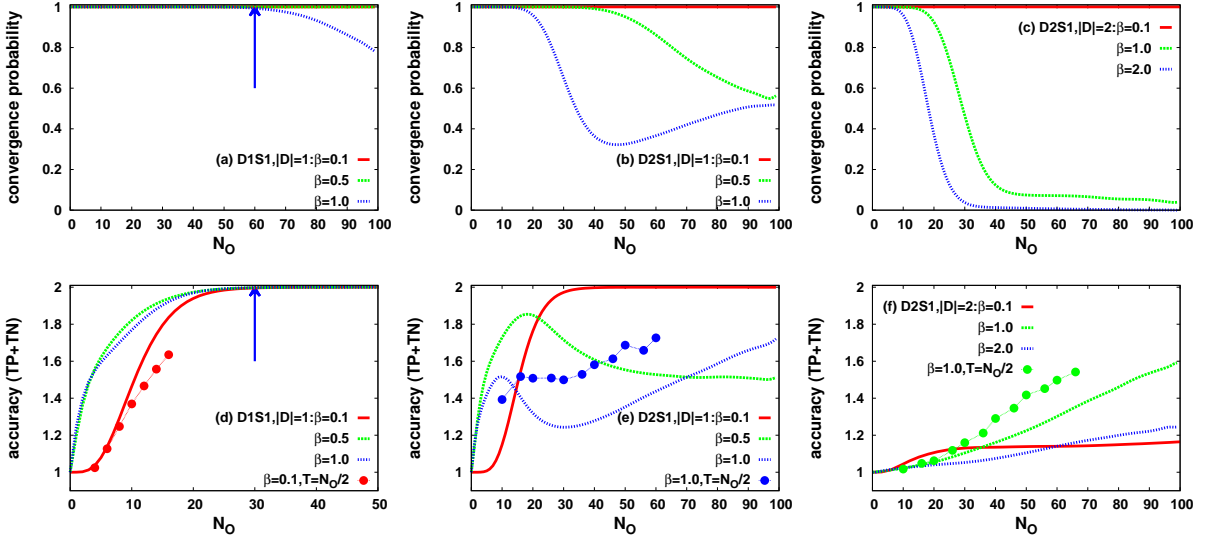} 
\caption{Comparing the convergence probability (panels a,b,c) and accuracy (true positive plus true negative) of the diagnosis with and without extrapolation (panels d,e,f) for the D1S1 and D2S1 models using the MF approximation. We consider the cases in which only one or two diseases are present $(|\mathbf{D}|=1,2)$. The prior disease probabilities are chosen such that $N_DP_0(D_a=1)=|\mathbf{D}|$. The model parameters are obtained from the exponential true model. The filled circles show the results after a simulation process of $T=N_O/2$ steps, where $N_O$ is the initial number of the observed signs with known true values. For the model structure we take a random graph of $N_D=50, N_S=500$ variables with $M_a=50, M_{ab}=500$ interaction factors, and connectivities $k_a=k_{ab}=400$. The data are results from at least $100$ independent realizations of the problem.}\label{MF}
\end{figure}

Suppose that we are given a subset $\mathbf{O}$ of $N_O$ observed signs with values $\mathbf{S}^o$. A simple diagnostic procedure works by computing the posterior disease probabilities conditioned on the observations $P(D_a|\mathbf{S}^o)$. Then, the most probable diseases or those that have a probability greater than a threshold value, are reported as the diagnosed diseases; in the following, we shall assume that the most probable diseases, within a small window of size $\delta P_D=0.01$, are the present ones. Figure \ref{EX} displays the accuracy of such a diagnosis with the D1S1 and D2S1 models for a small number of sign and disease variables. The figure also shows the probability gap between the most probable disease(s) and the other diseases. A patient with disease pattern $\mathbf{D}$ and $N_O$ initial observed signs from the most probable symptoms $\mathbf{S}(\mathbf{D})$ is presented to the model for diagnosis; a disease pattern is chosen with a probability proportional to the weights $W_a$ of the present diseases in $\mathbf{D}$. From \cite{paper-1} we know that the D1S1 and D2S1 models work well so long as the number of present diseases in $\mathbf{D}$, denoted by $|\mathbf{D}|$, is less than or equal to two; that is why we choose patients with a small number of diseases. As the figure shows, we obtain more accurate predictions as the parameter $\beta$ increases. The situation is different when we have to resort to an approximate inference algorithm. We see in Fig. \ref{MF} that the MF approximation does not provide accurate estimations of the sign and disease marginal probabilities for large $\beta$, where the algorithm does not converge. Here the best performances are observed for intermediate values of $\beta$.

\section{Diagnosis by simulation of the diagnostic process}\label{S4}
It may happen that the information provided by the initial number of observations are not enough to reach a reliable diagnosis, especially in the early stages of the diseases. Thus, we need a good strategy to choose the most informative signs for the next observations. Here the goal could be to reach the right diagnosis with a minimum number of the medical tests \cite{paper-1,paper-2}. Thus, for the objective function we propose an increasing function of the polarizations (deviations from the neutral value) in the posterior disease probabilities. The optimal choice then is provided by the most polarizing observation conditioned on the value of the previous observations. In contrary to the maximum likelihood function which is computationally hard to compute, the above objective function can easily be computed given the posterior disease probabilities. And, one can easily incorporate the importances of the diseases (the $W_a$) into the objective function, to assign more weight to polarization of the more important diseases. 

In a sequential diagnostic process of length $T$, we do the medical tests one by one and at each step we obtain the true value of the observed sign (this is called Diags-I in \cite{paper-2}). To obtain an optimal sequence of medical tests, one has to simulate in advance a diagnostic process of $T$ observations without doing any real observation (this is called Diags-II in \cite{paper-2}). This simulation of the observation process, or extrapolation from the initial set of observations, is proposed here as another heuristic approach to diagnosis to fully exploit the statistical dependencies of the variables provided by the model in addition to the initial medical tests. To this end, we use the mean-field approximation to compute the posterior disease probabilities. Then, the (Monte Carlo) optimization algorithm of Ref. \cite{paper-2} is used to maximize an objective functional of $T$ observations, 
\begin{align}\label{E}
\mathcal{E}[\mathbf{O}(T)]= \sum_{t=1}^T \left(\frac{1}{\sum_a W_a}\sum_a W_a \left| P(D_a=1)-\frac{1}{2}\right| \right).
\end{align}
Here $\mathbf{O}(T)=\{j_1,\dots,j_T\}$ is the sequence of observations. One can also add the cost or relevance of the observed signs to this objective function \cite{paper-2}. The optimization algorithm starts from a random sequence of $T$ observations, uses the marginal sign probabilities to generate a sequence of new observations, and accepts the suggested sequence if the objective function increases.    

Note that the above problem is indeed a stochastic optimization problem, where the objective function depends on the stochastic outcomes of the observations \cite{BL-book-1997,ABRZ-prl-2011}. To simplify the computation, we assume that the observed sign $j$ at each time step takes the most probable value identified by the marginal probability $P(S_j|\mathbf{S}^o)$ conditioned on the value of the previous observed signs.  

\begin{figure}
\includegraphics[width=16cm]{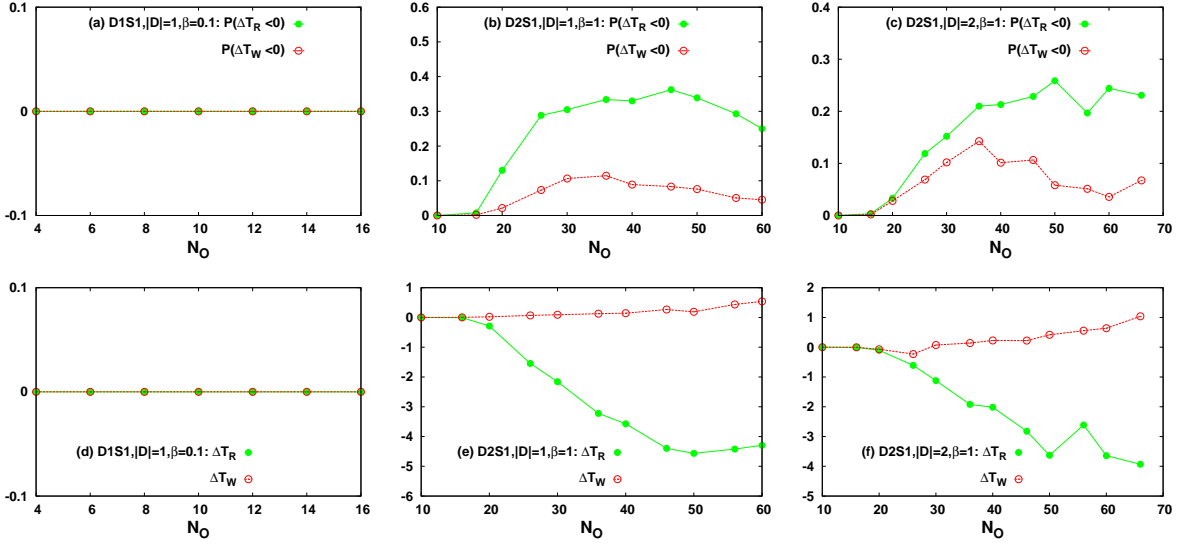} 
\caption{The improvment in the statistics of the first right and wrong diagnosis times $(T_R, T_W)$ after maximizing the objective function $\mathcal{E}[\mathbf{O}(T)]$ with the D1S1 and D2S1 models using the MF approximation. The algorithm starts from a random sequence of $T=N_O/2$ observations, where $N_O$ is the initial number of the observed signs with known true values. Panels (a,b,c) show the probabilities $P(\Delta T_{R,W}<0)$ of decreasing the corresponding times by the algorithm. Panels (d,e,f) show the average values $\Delta T_{R;W}$ of the changes in the corresponding times by the algorithm. We consider the cases in which only one or two diseases are present $(|\mathbf{D}|=1,2)$. The prior disease probabilities are chosen such that $N_DP_0(D_a=1)=|\mathbf{D}|$. The model parameters are obtained from the exponential true model. For the model structure we take a random graph of $N_D=50, N_S=500$ variables with $M_a=50, M_{ab}=500$ interaction factors, and connectivities $k_a=k_{ab}=400$. The data are results of at lesat $100$ independent realizations of the problem.}\label{MF-dT}
\end{figure}

\begin{figure}
\includegraphics[width=10.5cm]{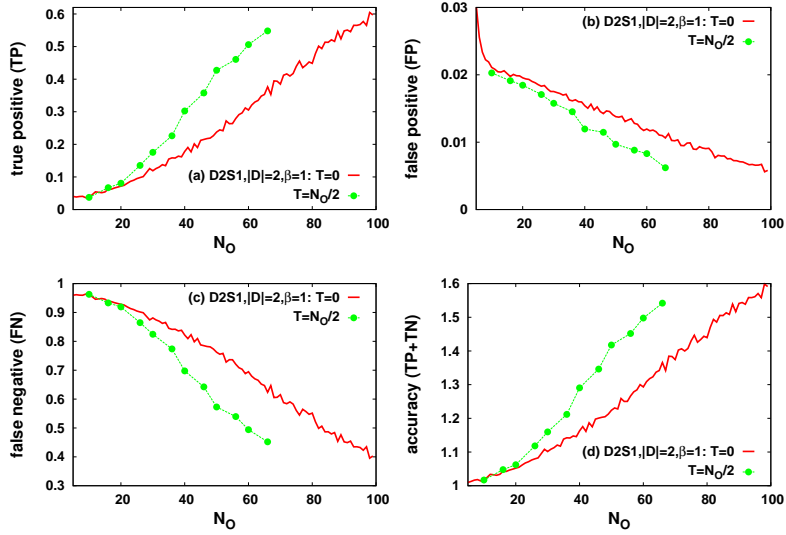} 
\caption{Comparing the accuracy of the diagnosis with and without extrapolation for the D2S1 model using the MF approximation. We consider the cases in which only two diseases are present $(|\mathbf{D}|=2)$. The prior disease probabilities are chosen such that $N_DP_0(D_a=1)=|\mathbf{D}|$. The model parameters are obtained from the exponential true model. The filled circles show the results after a simulation process of $T=N_O/2$ steps, where $N_O$ is the initial number of the observed signs with known true values. For the model structure we take a random graph of $N_D=50, N_S=500$ variables with $M_a=50, M_{ab}=500$ interaction factors, and connectivities $k_a=k_{ab}=400$. The data are results of at lesat $100$ independent realizations of the problem.}\label{MF-D2S1}
\end{figure}

Figure \ref{MF-dT} shows how the above objective function and the optimization algorithm perform. The figure displays the changes the first right and wrong diagnosis times compared to a random sequence of $T$ observations  \cite{paper-2}; the first right diagnosis time $T_R$ is the first time (number of observations) the probability of having a right disease becomes larger than a threshold value, here $P_{th}=0.9$. Similarly we define the first wrong diagnosis time $T_W$. In Fig. \ref{MF-D2S1}, we compare the accuracy of the diagnosis with the D2S1 model before and after extrapolation for $T=N_O/2$ steps. Here the sign/disease marginal probabilities are computed by the MF approximation. Similar comparisons are shown also in Figs. \ref{EX} and \ref{MF}.

\section{Discussion}\label{S5}
In summary, depending on the model and the strength of the model parameters, new macroscopic states can appear as the number of observed signs increases. This could be helpful because the disease probabilities are usually more informative within such states. On the other hand, this affects the algorithm convergence and consequently the quality of the sign and disease probabilities which are computed by the approximate inference algorithm. More advanced and accurate algorithms can of course improve the quality of inference, but at the expense of more computational time \cite{paper-1}.    

We showed that simulation of the diagnostic process provides a useful strategy for diagnosis when a naive approach that is based on the direct implications of the observed signs is not very helpful. In other words, this strategy works in the ordered phase of the system where the values of the observed signs significantly affect the probability distribution of the unobserved signs; the classical example is a ferromagnetic spin system in the ordered (low temperature) phase where the values of the boundary spins determine the physical (Gibbs) state of the system. In this way, we can define a critical number of initial observations which are needed to enter such an ordered state, for systems that display a phase transition.   

Here, for the sake of efficiency, we assumed that each "observed" sign in the simulation takes the most probable value predicted by the model. Moreover, we used a very naive optimization algorithm to find the optimal sequence of observations. A more accurate study should consider the stochastic nature of the simulated observations, and employ a more sophisticated optimization algorithm, e.g., simulated annealing. Finally, it would be interesting to have a microscopic (or phenomenological) model of patient (or an ensemble of patients) to study the time evolution of the sign probability distribution, and the emergent macroscopic (disease) states.           

\section{Perspectives}\label{S6}
An accurate medical diagnosis from a limited number of findings (e.g. at the early stages of diseases) should exploit all the statistical information on the sign/disease dependencies observed in the clinical and laboratory data. Such interdependencies are emerging due to the advancements in omics technologies and progress in population studies and aging research (e.g. identification of co-occurrence of age-related diseases). We note that the existing datasets lack the necessary probabilistic information needed for our approach, as such new data need to be generated. These studies will be the subject of our future works and in the current article, we are primarily addressing the mathematics and statistical physics communities.

Let us recall briefly the kind of statistical data we need to construct the models studied in this paper. First, note that these models have been obtained from an expansion around the healthy state where the number of involved diseases is small ($|\mathbf{D}|=1,2$) \cite{paper-1}. On the other hand, given a disease hypothesis $\mathbf{D}$, it is usually assumed that the sign variables are uncorrelated in a zero-order approximation of the signs statistics \cite{internist-1}. Here, the necessary data are encoded in the conditional probabilities $P(S_i|\mathrm{only}D_a)$ (in D1S1 model) or $P(S_i|\mathrm{only}D_a,D_b)$ (in D2S1 model). Obviously, we expect to have two-sign correlations, or higher-order sign correlations, even in presence of only a single disease. But taking into account these correlations considerably increases the computational complexity of the problem. Additionally, it is in practice very difficult to obtain statistically good clinical data which capture the higher-order correlations. Nevertheless, in the end, it is the collection of available empirical data that determines the structure of the model. 

The method can in principle be applied to any diagnostic problem to infer the macroscopic state (phenotype) of the system from a limited number of evidences. This could be, for instance, the problem of assigning a state to a biological cell or a complex electronic device. In particular, assignment of state to a cell is a major challenge in immunology and cancer biology and it has complicated developing therapies for cancer and autoimmunity. We envision that our approach will be generically applied to a wide range of problems in medicine, science and technology.

%\acknowledgments

\appendix

\section{The homogeneous fully-connected models}\label{app-1}
As long as the model parameters do not change with the sign or disease labels, we can write all the quantities in terms of the collective variables $x=(\sum_aD_a)/N_D$ and $y=(\sum_iS_i)/N_S$. Then for large number of signs ($N_S\to \infty$), we get
\begin{align}
\frac{1}{N_S}\ln Z(\mathbf{D}) &\approx \mathcal{S}(\frac{1+z(x)}{2})+h(x)z(x)+\frac{1}{2}J(x)z^2(x),\\
\frac{1}{N_S}\mathcal{H}(\mathbf{S};\mathbf{D}) &\approx -h(x)(y-z(x))-\frac{1}{2}J(x)(y^2-z^2(x))+\mathcal{S}(\frac{1+z(x)}{2})-\gamma \kappa_a^0 x,
\end{align}
where $\gamma=N_D/N_S$, and
\begin{align}
h(x) &=\kappa_i^0+\kappa_i^a x+\frac{1}{2}\kappa_i^{ab}x^2,\\
J(x) &=\kappa_{ij}^a x+\frac{1}{2}\kappa_{ij}^{ab}x^2.
\end{align}
Here we take the scaling 
\begin{align}
K_i^0 &=\kappa_i^0,\hskip0.5cm  K_a^0=\kappa_a^0,\\
K_i^a &=\frac{1}{N_D}\kappa_i^a,\hskip0.5cm K_i^{ab}=\frac{1}{N_D^2}\kappa_i^{ab},\\
K_{ij}^a &=\frac{1}{N_SN_D}\kappa_{ij}^a,\hskip0.5cm K_{ij}^{ab}=\frac{1}{N_SN_D^2}\kappa_{ij}^{ab}.
\end{align}
Moreover, $z$ is the solution to
\begin{align}
z=\tanh(h(x)+J(x) z).
\end{align}
which minimizes the following free energy
\begin{align}
f(z) = -\mathcal{S}(\frac{1+z}{2})-h(x)z-\frac{1}{2}J(x)z^2,
\end{align}  
Here, for brevity, we defined the Shanon entropy function
\begin{align}
\mathcal{S}(p)=-p\ln p-(1-p)\ln(1-p).
\end{align}

To take into account the value of the observed signs, we write $y=n_o y_o+(1-n_o)y_u$ with $y_o=(\sum_{i\in \mathbf{O}}S_i^o)/N_O$, $y_u=(\sum_{i\notin \mathbf{O}}S_i)/(N_S-N_O)$, and $n_o=N_O/N_S$. In this way, the grand partition function is given by
\begin{align}
\mathcal{Z}(\mathbf{S}^o) \simeq \int_0^1 dx \int_{-1}^1 dy_u e^{-N_S f(x,y)},
\end{align}
At the end, the self-consistency equations for the $x$ and $y_u$ variables in the thermodynamic limit ($N_S\to \infty$), are obtained by minimizing the following free energy 
\begin{multline}
f(x,y)=-\gamma \mathcal{S}(x)-(1-n_o)\mathcal{S}(\frac{1+y_u}{2})\\
-h(x)(y-z(x))-\frac{1}{2}J(x)(y^2-z^2(x))+\mathcal{S}(\frac{1+z(x)}{2})-\gamma \kappa_a^0 x.
\end{multline}
Figure \ref{HMF-fx} shows how the above free energy behaves when the model parameters in the D2S1 model are varied.

\section{The inhomegeuous models: mean-field approximation}\label{app-2}
For the D1S1 and D2S1 models we can compute some quantities exactly, therefore, we present the mean-field approximation for these models separately.
 
\subsection{In the absence of the sign-sign interactions}\label{app-21}
Here we write $D_a=x_a+\delta D_a$ where $x_a=\langle D_a \rangle$, and $\delta D_a=D_a-x_a$ is a small deviation from the mean value.
Then, the local field experienced by sign $i$ is
\begin{align}
h_i(\mathbf{D})= h_i(\mathbf{x})+ \sum_a K_i^a \delta D_a+\sum_{a<b}K_{i}^{ab}(\delta D_ax_b+x_a\delta D_b+\delta D_a\delta D_b)=h_i(\mathbf{x})+\delta h_i,
\end{align}
where
\begin{align}
h_i(\mathbf{x}) &= K_i^0+ \sum_a K_i^a x_a+\sum_{a<b}K_{i}^{ab}x_ax_b.
\end{align}
The Hamiltonian can be written as
\begin{align}
\mathcal{H}(\mathbf{D}|\mathbf{S}^o) =  -\sum_a K_a^0D_a-
\sum_{i\in \mathbf{O}}S_i^o[h_i(\mathbf{x})+\delta h_i]+\sum_{i\in \mathbf{O}}\ln \left(2\cosh [h_i(\mathbf{x})+\delta h_i] \right),
\end{align}
Expanding the last term up to the first order deviations $\delta D_a$, we get
\begin{align}
\mathcal{H}(\mathbf{D}|\mathbf{S}^o) \approx \mathcal{H}_0-\sum_a h_a(\mathbf{x}) D_a,
\end{align}
with
\begin{align}
h_a(\mathbf{x}) &= K_a^0+\sum_{i\in \mathbf{O}}[S_i^o-\tanh(h_i(\mathbf{x}))](K_i^a+\sum_{b\neq a}K_i^{ab}x_b).
\end{align}

Then, the average values $x_a$ are obtained by the following self-consistency equations:
\begin{align}
x_a =\frac{e^{h_a(\mathbf{x})}}{1+e^{h_a(\mathbf{x})}},
\end{align}
The equations are solved by iteration starting from random initial values for the $x_a$.

\subsection{In the presence of the sign-sign interactions}\label{app-22}
In general, the partition function in the MF approximation reads
\begin{align}
Z(\mathbf{D}) \propto \prod_i \left(2\cosh[h_i(\mathbf{D})+\sum_{j\neq i}J_{ij}(\mathbf{D})z_j(\mathbf{D})] \right),
\end{align}
where the $z_i$ are solutions to
\begin{align}
z_i(\mathbf{D})=\tanh[h_i(\mathbf{D})+\sum_{j\neq i}J_{ij}(\mathbf{D})z_j(\mathbf{D})].
\end{align}
Define $\delta D_a=D_a-\langle D_a \rangle$ and $\delta S_i=S_i-\langle S_i \rangle$. For brevity, we take $x_a=\langle D_a \rangle$ and $y_i=\langle S_i \rangle$.
Note that $y_i=S_i^o$ is fixed for $i\in \mathbf{O}$ (the subset of observed signs).

Then, to first order in the $\delta D_a$ and $\delta S_i$, we have
\begin{align}
\mathcal{H}(\mathbf{S};\mathbf{D}) \approx \mathcal{H}_0 -\sum_{i\notin \mathbf{O}} [h_i(\mathbf{x})+\sum_{j\neq i}J_{ij}(\mathbf{x})y_j] S_i-\sum_{a} h_a(\mathbf{x},\mathbf{y},\mathbf{z}) D_a,
\end{align}
where
\begin{multline}
h_a(\mathbf{x},\mathbf{y},\mathbf{z})=K_a^0+\sum_i B_i^a(\mathbf{x}) y_i+\sum_{i<j} B_{ij}^a(\mathbf{x}) y_iy_j \\
-\sum_i\tanh[h_i(\mathbf{x})+\sum_{j\neq i}J_{ij}(\mathbf{x})z_j]
\times \left( B_i^a(\mathbf{x})+\sum_{j\neq i}[B_{ij}^a(\mathbf{x})z_j+J_{ij}(\mathbf{x})\chi_j^a]\right) 
\end{multline}
Here, the new introduced local fields are
\begin{align}
B_i^a(\mathbf{x}) &=K_i^a+\sum_{b\neq a}K_i^{ab}x_b,\\
B_{ij}^a(\mathbf{x}) &=K_{ij}^a+\sum_{b\neq a}K_{ij}^{ab}x_b,
\end{align}
and the susceptibility $\chi_i^a$ is given by 
\begin{multline}
\chi_i^a = \frac{\partial z_i}{\partial x_a}=(1+\tanh^2[h_i(\mathbf{x})+\sum_{j\neq i}J_{ij}(\mathbf{x})z_j])\\
\times\left( B_i^a(\mathbf{x})+\sum_{j\neq i}[B_{ij}^a(\mathbf{x})z_j+J_{ij}(\mathbf{x})\chi_j^a]\right).
\end{multline}

In summary, the mean-field equations read as follows,
\begin{align}
x_a &=\frac{e^{h_a(\mathbf{x},\mathbf{y},\mathbf{z})}}{1+e^{h_a(\mathbf{x},\mathbf{y},\mathbf{z})}},\\
y_i &=S_i^o,\hskip0.5cm i\in \mathbf{O}\\
y_i &=\tanh[h_i(\mathbf{x})+\sum_{j\neq i}J_{ij}(\mathbf{x})y_j],\hskip0.5cm i\notin \mathbf{O}\\
z_i &=\tanh[h_i(\mathbf{x})+\sum_{j\neq i}J_{ij}(\mathbf{x})z_j].
\end{align}
We solve the equations by iteration starting from random initial values for the $x_a$, $y_i (i\in \mathbf{O}),z_i$, and the $\chi_i^a$.

%%%%%%%%%%%%%%%%%%%%%%%%%%%%%%%%%%%%%%%%%%% main figures

%%%%%%%%%%%%%%%%%%%%%%%%%%%%%%%%%%%%%%%%%%% app figures

\begin{figure} 
\includegraphics[width=10.5 cm]{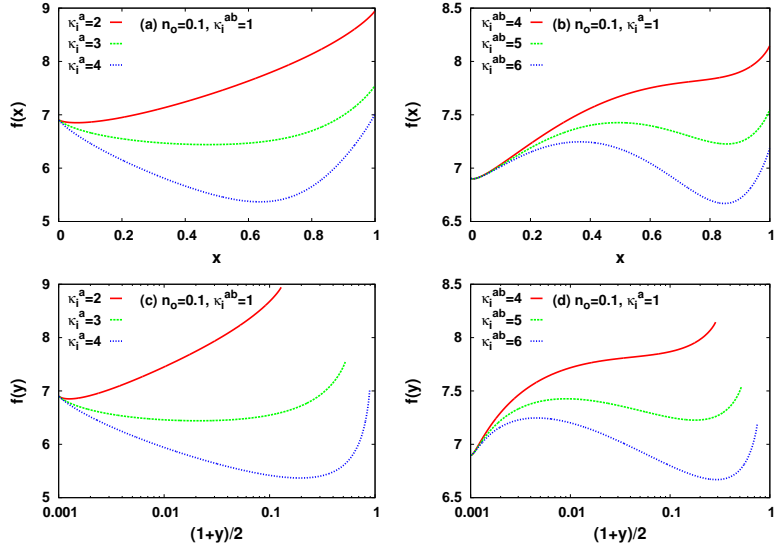} 
\caption{The free energy landscape as a function of the sign and disease probabilities in the homogenuous fully-connected D2S1 model. We assume that all the observed signs are positive. Here $P(D=1)=x$, $P(S=+1)=(1+y)/2$, and $n_o=N_O/N_S$. The prior disease probabilities and the leak sign probabilities are $P_0(D_a=1)=P(S_i=+1|\mathbf{0})=0.001$.}\label{HMF-fx}
\end{figure}

\end{document}